\documentclass{eptcs}
\usepackage{latexsym,amssymb}
\usepackage[leqno,fleqn]{amsmath}
\usepackage{proof}
\newtheorem{defn}{Definition}
\newtheorem{thm}[defn]{Theorem}

\newtheorem{lem}[defn]{Lemma}
\newcommand{\proofbeg}{\textbf{Proof. }}
\newcommand{\proofend}{\hfill $\square$}

\newcommand{\KA}{\mathcal{K}}
\newcommand{\osigma}{\overline{\sigma}}
\newcommand{\KL}{\mathsf{KL}}

\title{Kleene Algebras, Regular Languages and Substructural Logics}
\author{Christian Wurm \\
\texttt{cwurm@phil.uni-duesseldorf.de}}
\def\titlerunning{Kleene Algebras, Regular Languages and Substructural Logics}
\def\authorrunning{Christian Wurm}
\begin{document}

\maketitle
\begin{abstract}
We introduce the two substructural propositional logics $\KL$, $\KL^+$, 
which use disjunction, fusion and a unary, (quasi-)exponential 
connective. For both we prove strong completeness
with respect to the interpretation in Kleene 
algebras and a variant thereof.
We also prove strong completeness  for language models, where each
logic comes with a different interpretation.
We show that for both logics the cut rule is 
admissible and both have a decidable consequence relation.
\end{abstract}

\section{Introduction and Motivation}

We introduce two substructural logics, the logic $\KL$ and
the logic $\KL^+$, by providing a Gentzen-style sequent calculus.
$\KL$ and $\KL^+$ have the same syntax: they are propositional, 
and thus consist of a countable set of propositional variables together with
propositional connectives. They have the binary
connectives $\vee,\bullet$, where $\vee$ is the classical \textit{or},
and $\bullet$ is the \textit{fusion} operator well-known from 
Lambek calculus, which is non-commutative and non-monotonic in
both directions. Moreover, they have the unary connective $?$, which is similar
to the right exponential from linear logic, in the sense that it 
allows weakening and contraction on the right hand side of $\vdash$
and can be introduced only under strict conditions on the left hand 
side.\footnote{See \cite{galatos:residuated} and \cite{restall:introduction} 
for treatment of the full Lambek calculus; \cite{girard:linear} for 
linear logic.}

We show that $\KL$ can be interpreted in Kleene algebras (we use this
term in the sense of the axiomatization of \cite{kozen:kleene}) in a
natural way, and that this results in a strongly complete semantics
We can thus interpret formulas as languages with the standard interpretation
of regular expressions, consequence as set-theoretic inclusion,
and still have a strongly complete semantics. 
A similar approach has be taken by \cite{kozen:substructural}, but our
calculus differs substantially: in particular, it has a pure
Gentzen-style presentation, no structural rules, and apart from the (admissible) 
cut rule, its syntactic decidability can be neatly read off.
$\KL^+$ differs minimally, but its semantics is much
less obvious: $\KL^+$ is a fragment of $\KL$, that is, every
valid sequent of $\KL^+$ is valid in $\KL$, but not vice versa. 
We show that
we can interpret it in a closely related variant of Kleene algebras,
and what is more interesting: we can interpret  expressions of these
algebras (and thus $\KL^+$ formulas) 
as languages in the usual sense, with the only difference
that $?$ is interpreted as Kleene plus instead of star, and get a
strongly complete language-theoretic semantics. From this we can easily
conclude that both logics are decidable (because the inclusion problem
for regular languages is decidable), and that the cut rule
is admissible for both logics: for every provable sequent, there is a
cut-free proof.

There are two main motivations for $\KL$: as it is sound and complete
for Kleene algebras, we hardly need to explain its many possible
interpretations. What is firstly interesting about $\KL$ is that it connects
logical questions with language-theoretic questions and regular expressions
(as has been done, in a rather different setting, by
\cite{babu:chop}). We thus can use language-theoretic
techniques to check theoremhood and consequence; and conversely, we can
use our cut-free sequent calculus to check whether the language denoted
by one regular expression is a subset of a language denoted by another 
one (which is still a ``hot topic" in computer science, see e.g.
\cite{hovland:inclusion}).
The main motivation for $\KL^+$ is somewhat philosophical, but also
has a computational content: if we want to
interpret $\KL$ as a logic of events/processes and $?$ as a \textit{progressive
aspect} of a process (saying it is going on in some interval
rather than being completed in this interval), 
some of the $\KL$ rules seem to be too strong,
whereas those of $\KL^+$ seem to be reasonable. It is interesting and
instructive that many nice results from $\KL$ -- cut-free sequent
calculus, decidable and complete algebraic and language-theoretic
semantics -- can be transferred to $\KL^+$; in particular, the
language-theoretic semantics of the latter is far from obvious.

\section{Syntax and Sequent Calculus of $\KL$}

We now present the syntax of both $\KL$ and $\KL^+$. 
 The set of formulas is defined as follows:
let $\textit{Var}$ be a countably infinite set of variables. Then
we define:
\begin{enumerate}
\item If $\alpha\in \textit{Var}$, then $\alpha\in \textit{Form}(\textit{Var})$.
\item If $\alpha\in \textit{Form}(\textit{Var})$, then $\alpha?\in \textit{Form}(\textit{Var})$.
\item If $\alpha,\beta\in \textit{Form}(\textit{Var})$, then $\alpha\bullet\beta\in \textit{Form}(\textit{Var})$.
\item If $\alpha,\beta\in \textit{Form}(\textit{Var})$, then $\alpha\vee\beta\in \textit{Form}(\textit{Var})$.
\end{enumerate}

This defines the set of formulas. I first say a word on the
intuitive meaning of formulas. Atomic propositions might be best
thought of as events or actions. The $\vee$ should be clear,
representing a classical ``or" in a non-classical context. The $\bullet$
can be read as ``and then", meaning temporal sequence of events. The
intuitive meaning of $?$ can be thought of: ``is happening (or taking place)
some arbitrary number (including 0) of times". This is very close
to the intuitive interpretation of operations in Kleene algebras;
for more explicit considerations consider \cite{pratt:action}.

We let lowercase Greek letters range over formulas, uppercase
Greek letters over finite, possibly empty sequences of 
formulas,
which we write in the usual fashion just separated by
$``,"$. $\KL,\KL^+$ are substructural logics, that is, the usual
structural rules of weakening, contraction, monotonicity are
not legitimate. If we
present a sequence of formulas separated by $``,"$, this means that the
sequence is ordered: we cannot exchange neither order nor 
cardinality of its elements, nor can we add or take away anything without an
additional rule. The
$``,"$ is however associative, that is, 
for a sequence $\alpha,\beta,\gamma$
there is no precedence.
Now we define the derivability relation 
$\vdash$ of the logic $\KL$ between sequences of formulas and
formulas. A substructural consequence relation as the one we
present here also gives rise to a structural consequence
relation (see \cite{galatos:residuated}), but we will not be
concerned with this here.
\\

(ax) $\infer{\alpha\vdash\alpha}{}$
$\qquad\qquad$ 
(cut) $\infer{\Gamma,\Theta,\Delta\vdash \beta}
{\Gamma,\alpha,\Delta\vdash \beta & \Theta\vdash\alpha}$
\\

($\vee$I) $\infer{\Gamma,\alpha\vee\beta,\Delta\vdash\gamma}
{\Gamma,\alpha,\Delta\vdash \gamma & \Gamma,\beta,\Delta\vdash\gamma}$
$\qquad\qquad$ (I$\vee$1) $\infer{\Gamma\vdash \alpha\vee\beta}{\Gamma\vdash \alpha}$
$\qquad\qquad$
(I$\vee$2) $\infer{\Gamma\vdash \alpha\vee\beta}{\Gamma\vdash \beta}$
\\

(I$\bullet$) $\infer{\Gamma,\Delta\vdash \alpha\bullet\beta}
{\Gamma\vdash \alpha & \Delta\vdash \beta}$
$\qquad\qquad$
($\bullet$I) $\infer{\Gamma,\alpha\bullet\beta,\Delta \vdash \gamma}
{\Gamma,\alpha,\beta,\Delta\vdash \gamma}$
\\

These rules are the usual axioms of sequent calculi, and usual rules for
$\vee,\bullet$ which should be familiar to anyone familiar with some 
substructural logic. In addition, we have to make sure that our logic
satisfies the distributive law for $\vee$ and $\bullet$: 
(DIS1) $(\alpha\bullet\beta)\vee(\alpha\bullet\gamma)
\vdash\alpha\bullet(\beta\vee\gamma)$
and
(DIS2) $\alpha,\beta\vee\gamma\vdash (\alpha\bullet\beta)\vee(\alpha\bullet\gamma)$.
(DIS1) is derivable from
the above rules, but (DIS2) is not (see \cite{restall:introduction}
for background), so 
we have to add a rule to make sure (DIS2) holds:
\\

(D) $\infer{\Gamma\vdash (\alpha\bullet\beta)\vee(\alpha\bullet\gamma)}
{\Gamma\vdash \alpha\bullet(\beta\vee\gamma)}$
\\

This way, we make sure the distributive law holds for our logic.
Now comes the most interesting group of axioms and rules, namely the ones for
$?$. 
\\

(ax?) $\infer{\vdash \alpha?}{}$
$\qquad\qquad$
(I?1) $\infer{\Delta,\Gamma\vdash \alpha?}
{\Delta\vdash\alpha & \Gamma\vdash \alpha?}$
$\qquad\qquad$
(I?2) $\infer{\Gamma,\Delta\vdash \alpha?}
{\Delta\vdash\alpha & \Gamma\vdash \alpha?}$
\\

(I?1),(I?2) form a symmetric group; we use 
it in order to introduce formulas to the
left of the turnstile, if on the right there is a formula $\alpha?$. By 
(ax?), the problem is not to get a formula $\alpha?$ 
on the right, but rather introduce
material on its left. The next group is responsible for introducing
the $?$ on the left side of the turnstile.
\\

(?I1) $\infer{\alpha?,\Gamma\vdash\beta}
{\alpha,\beta\vdash\beta & \Gamma\vdash\beta}$
$\qquad\qquad$
(?I2) $\infer{\Gamma,\alpha?\vdash\beta}
{\beta,\alpha\vdash\beta & \Gamma\vdash\beta}$
\\

This is sufficient for $?$
We now introduce the constants $1,0$ as follows; note that we can also do
without, and the extension we thereby introduce is conservative.
\\

(1I) $\infer{\Gamma,1,\Delta\vdash\alpha}{\Gamma,\Delta\vdash\alpha}$
$\qquad\qquad$
(I1) $\infer{\vdash 1}{}$
$\qquad\qquad$
(0I) $\infer{\Gamma,0,\Delta\vdash\alpha}{}$
\\

Note that our 1 is weak, in the sense that it is not necessarily
derivable from any sequence; our 0 on the other side is strong:
everything is derivable from any sequence containing it.
So the two are not duals, and we do not have a strong 1 (usually 
written $\top$)
or a weak 0 (the strong 0 is usually denoted by $\perp$).
From the axioms it follows that 1 is the ``neutral element" 
(we already speak in algebraic terms) of $\bullet$, and
$0$ the ``neutral element" of $\vee$. Moreover, $0?$ is logically
equivalent to
$1$, and $1?$ equivalent to $1$. The cut rule goes without comment,
but we already presage that we will later show that it is admissible,
that is, anything derivable with cut is also derivable without it.
We define a $\KL$-\textbf{proof tree}
as usual in logic: it is a (binary) labelled tree where each leaf is
labelled by an axiom, and each elementary subtree (node with its
two daughters) is labelled according to one of our inference rules. 
We say a sequent $\Gamma\vdash\alpha$ is derivable (in $\KL$) by our
proof calculus, if there is a $\KL$-proof tree such that
its root is labelled by $\Gamma\vdash\alpha$.
If such a sequent is derivable, 
we write $\Vdash_{\KL}\Gamma\vdash\alpha$.

\section{Kleene Algebras, Semantics of $\KL$ and its Completeness}

For the semantics of $\KL$ we use a well-known class of structures
(see \cite{kozen:kleene}, and \cite{conway:regular} as classical reference).
A \textbf{Kleene algebra} is an algebra $(K,+,\cdot,*,0,1)$, 
where $+,\cdot$ are binary operators, $*$ is unary, $0,1$ are
constants in $K$ and
$K$ is a set closed under the former
operations, and satisfying the following equations:
\\

(K1) $(a+b)+c=a+(b+c)$

(K2) $a+b=b+a$

(K3) $a+a=a$

(K4) $a+0=a$

(K5) $a\cdot(b\cdot c)=(a\cdot b)\cdot c$

(K6) $a\cdot 1=1\cdot a=a$

(K7) $0\cdot a=a\cdot 0=0$

(K8) $a\cdot(b+c)=a\cdot b+a\cdot c$

(K9) $(b+c)\cdot a=b\cdot a+c\cdot a$
\\

This means: $(K,+,\cdot,0,1)$ is an idempotent semiring. We write $ab$ for
$a\cdot b$; as usual, $\cdot$ has precedence over $+$. 
 We also define a partial order
on $K$ as usual: $a\leq b$ if and only if $a+b=b$.
To get a Kleene algebra, we still need two inequations and two 
quasi-equations for $*$:
\\

(K10) $1+aa^*\leq a^*$

(K11) $1+a^*a\leq a^*$

(K12) If $b+ac\leq c$, then $a^*b\leq c$

(K13) If $b+ca\leq c$, then $ba^*\leq c$.
\\

These axioms say that $a^*b$ is the unique smallest solution for $x$
in the inequation $b+ax\leq x$. As usual, $*$ has precedence over 
both $+,\cdot$. By $\KA$ we denote the class of
all Kleene algebras, that is, all algebras satisfying 
(K1)--(K13). Let $\textbf{K}$ be a Kleene algebra. If an equation
(or inequation) $a=b$ ($a\leq b$) holds in $\textbf{K}$, we sometimes
write $a=_\textbf{K}b$ (or $a\leq_\textbf{K} b$) for reasons of clarity.
If a certain (in)equation is valid in all Kleene algebras, we write
$a=_\KA b$ ($a\leq_\KA b$) (equivalently, if it is valid in the free
Kleene algebra, see \cite{burris} for algebraic background).
 Two properties of Kleene algebras are extremely
important for us: 1. $\leq_\textbf{K}$ is transitive
for any Kleene algebra $\textbf{K}$ (and so is $\leq_\KA$), and secondly, $\leq_\KA$
respects concatenation: if 
$a\leq b,c\leq d$, then $ac\leq bd$. This follows from distributivity:
$bd=(a+b)(c+d)=a(c+d)+b(c+d)=ac+ad+bc+bd=ac+ac+ad+bc+bd$. This is 
equivalent to saying that from $a\leq b$ it follows that $cad\leq cbd$,
a property to which we refer as \textbf{strong transitivity}
(this is also known as monotonicity of $\cdot$, but we prefer to use
monotonicity in connection with weakening).

An \textbf{interpretation} 
$\overline{\sigma}:\textit{Form}(\textit{Var})\rightarrow\textbf{K}$ is defined as
follows: let $\sigma:\textit{Var}\rightarrow K$ be an arbitrary
map from variables to $K$; 
we obtain $\overline{\sigma}$ inductively as follows:
\begin{enumerate}
\item $\osigma(\alpha)=\sigma(\alpha)$, if $\alpha\in \textit{Var}$.
\item $\osigma(\alpha\vee\beta)=\osigma(\alpha)+\osigma(\beta)$
\item $\osigma(\alpha\bullet\beta)=\osigma(\alpha)\cdot\osigma(\beta)$
\item $\osigma(\alpha?)=(\osigma(\alpha))^*$
\item $\osigma(1)=1$
\item $\osigma(0)=0$
\end{enumerate}

We extend $\osigma$ also to \textit{sequences} of formulas:
for $\Gamma=\gamma_1,...,\gamma_i$, put 
$\osigma(\Gamma):=\osigma(\gamma_1\bullet...\bullet\gamma_i)$;
that is, the $``,"$ of sequences is interpreted in the same manner
as $``\bullet"$.
Having defined $\osigma$, we define a \textbf{model} as a tuple
$(\textbf{K},\sigma)$. We say a sequent $\Gamma\vdash\alpha$ of $\KL$ is
\textbf{true} in $(\textbf{K},\sigma)$, in symbols
$(\textbf{K},\sigma)\models\Gamma\vdash\alpha$, if
$\osigma(\Gamma)\leq \osigma(\alpha)$ holds in $\textbf{K}$
(we also write: $\osigma(\Gamma)\leq_\textbf{K} \osigma(\alpha)$).
In case we have a sequent of the form $\vdash\alpha$, where
$\Gamma$ is the empty sequence, we write
$(\textbf{K},\sigma)\models\ \vdash\alpha$ if
$1\leq_{\textbf{K}}\osigma(\alpha)$.

We write $\textbf{K}\models\Gamma\vdash\alpha$, if for all maps
$\sigma:\textit{Var}\rightarrow K$, we have
$(\textbf{K},\sigma)\models\Gamma\vdash\alpha$; and we write
$\models_{\KA}\Gamma\vdash\alpha$, if for all Kleene algebras
$\textbf{K}$ and valuations $\sigma:\textit{Var}\rightarrow K$, we have
$(\textbf{K},\sigma)\models\Gamma\vdash\alpha$. In that case, 
we also say that $\Gamma\vdash\alpha$ is \textbf{valid}.
Our first main theorem is the following:

\begin{thm}
$\Vdash_{\KL}\Gamma\vdash\alpha$ if and only if 
$\models_{\KA}\Gamma\vdash\alpha$.
\end{thm}

So we have soundness and strong completeness. We start
with the only if direction, which is the usual soundness.

\proofbeg
\textit{\textbf{Soundness}}: We perform the usual proof by induction; 
we only prove those
cases which are not standard and straightforward. 
(ax), ($\vee$I), (I$\vee$1),(I$\vee$2),
(I$\bullet$),($\bullet$I) are clear.
(D) follows from the distributive law in $\KA$.
\\

(ax?) should be clear: it follows immediately from (K10) 
that $1\leq_\KA a^*$ holds.
\\

(I?1) 
Assume $\osigma(\Delta)\leq_\KA \osigma(\alpha)$, and 
$\osigma(\Gamma)\leq_\KA(\osigma(\alpha))^*$. 
We have $aa^*\leq_\KA a^*$ (by (K13)), and $\leq_\KA$ is transitive and
respects concatenation. Consequently, 
$\osigma(\Delta)\osigma(\Gamma)\leq_\KA \osigma(\alpha)(\osigma(\alpha))^*
\leq_\KA (\osigma(\alpha))^*$.
\\

(I?2) is similar.
\\

(?I1) 
Assume $\osigma(\alpha)\osigma(\beta)\leq_\KA\osigma(\beta)$, and
$\osigma(\Gamma)\leq_\KA\osigma(\beta)$.
From the premises and the fact that from $a\leq_\KA c,b\leq_\KA c$ it follows
that $a+b\leq_\KA c$;\footnote{Because 
then $a+c=_\KA c=_\KA b+c=_\KA a+b+c$} we thus have
$\osigma(\Gamma)+\osigma(\alpha)\osigma(\beta)\leq_\KA \osigma(\beta)$,
and by (K12),
$\osigma(\alpha)^*\osigma(\Gamma)\leq_\KA \osigma(\beta)$, and thus
$\models_\KA \alpha?,\Gamma\vdash\beta$.
\\

(?I2) is similar.
\\



(1I) clear for (K6) and the fact that $``,"$ matches $\bullet$.

(I1) is also obviously correct, as the empty antecedent is interpreted as
1.

(0I) follows from (K7) and (K4). 

(cut) This corresponds to strong transitivity, which we have established for
$\KA$.
\\

This completes the soundness direction. 
\\

We now prove \textbf{\textit{completeness}}. We proceed in the usual fashion:
we construct the algebra $\textbf{T}$ of $\KL$-terms modulo $\KL$-equivalence, 
where $\leq_T$ is $\vdash_{\KL}$ modulo logical equivalence. We prove that 
$\textbf{T}$ is a Kleene algebra. From this we can conclude: if 
$\not\Vdash_{\KL}\Gamma\vdash\alpha$, then 
there is a Kleene algebra $\textbf{T}$, 
assignment $\sigma$, such  that 
$(\textbf{T},\sigma)\not\models\Gamma\vdash\alpha$,
and consequently, $\not\models_\KA\Gamma\vdash\alpha$. By
contraposition, it follows that if $\models_\KA\Gamma\vdash\alpha$,
then $\Vdash_{\KL}\Gamma\vdash\sigma$ (alternatively, this goes by a
direct argument: if something holds in all Kleene algebras, it holds
in $\textbf{T}$, therefore it is provable).

\textbf{Definition of $\textbf{T}$}, term algebra 
or Lindenbaum-Tarski construction.

Define $\sim_{\KL}\subseteq \textit{Form}(\textit{Var})\times \textit{Form}(\textit{Var})$ as follows:
$\alpha\sim_{\KL}\beta$ if $\Vdash_{\KL}\alpha\vdash\beta$ and
$\Vdash_{\KL}\beta\vdash\alpha$. This is obviously an equivalence 
relation, because $\vdash$ is reflexive and transitive. 
We now define $T:=\textit{Form}(\textit{Var})/\negmedspace\negmedspace\sim_{\KL}$, 
the set of equivalence classes of $\textit{Form}(\textit{Var})$ under 
$\sim_{\KL}$, which is the generator set of $\textbf{T}$. 
The operators and constants of our algebra are 
the constructors and constants of
$\textit{Form}(\textit{Var})$, so we get the algebra 
$\textbf{T}=(\textit{Form}(\textit{Var})/\negmedspace\negmedspace\sim_{\KL},\vee,\bullet,?,0,1)$, and for 
$a,b\in \textit{Form}(\textit{Var})/\negmedspace\negmedspace\sim_{\KL}$,
we define $a\leq_\textbf{T} b$ iff there are $\alpha\in a,
\beta\in b$, and $\Vdash_{\KL} \alpha\vdash \beta$; we define
the equality 
$\textbf{T}$ as usual: $a=_\textbf{T}b$, if $a\leq_\textbf{T} b$ 
and $b\leq_\textbf{T} a$. 

As usually, we have to show that $\sim_{\KL}$ is a congruence over
the constructors, in order to show that $\sim_\KL$ is a congruence
and thus the operations on equivalence classes are independent
of representatives. We indicate the usual inductive procedure:

1. If $a\sim_\KL b$, $c\sim_\KL d$, then $a\bullet c\sim_\KL b\bullet d$.
This is fairly straightforward:
\\

$\infer{a\bullet c\vdash b\bullet d}
{\infer{a,c\vdash b\bullet d}{a\vdash b& c\vdash d}}$, 
and the other way round.

2. If $a\sim_\KL b$, $c\sim_\KL d$, then $a\vee c\sim_\KL b\vee d$.
Straightforward:
\\

$\infer{a\vee c\vdash b\vee d}
{\infer{a\vdash b\vee d}{a\vdash b}&\infer{c\vdash b\vee d}{a\vdash d}}$,
and the other way round.

3. If $a\sim_\KL b$, then $a?\sim_KL b?$.

Take the following derivation:
\\

$\infer{a?\vdash b?}
{\infer{a,b?\vdash b?}{a\vdash b & b? \vdash b?}& \vdash b?}$
\\

So we have shown that $\sim_\KL$ is a congruence over constructors,
and thus the quotient algebra is well-formed and independent of
representatives. Therefore, we will proceed in the sequel as
if congruence classes of formulas were formulas, and not distinguish
the two notationally.
Now comes the crucial step, namely to show that $\textbf{T}$ actually
\textit{is} a Kleene algebra. We prove that by going through
the equations (K1)--(K13) one by one. Recall that for $\textbf{T}$,
the equations consist of two inequations; while in principle, we have
to prove both, we usually 
only prove one direction, as the other one is similar.
Also, we omit some proofs, which are well-known from
existing logics.
\\

(K1) $(a+b)+c=_{\textbf{T}}a+(b+c)$

Obvious, canonical proof.

(K2) $a+b=_{\textbf{T}}b+a$

Obvious, canonical proof.

(K3) $a+a=_{\textbf{T}}a$

Obvious.

(K4) $a+0=_{\textbf{T}}a$

$a\leq_\textbf{T} a\vee 0$ follows from (I$\vee$1),
$a\vee_\textbf{T} 0\leq_\textbf{T} a$ holds because 
$a\leq_\textbf{T} a,0\leq_\textbf{T} a$.

(K5) $a\cdot(b\cdot c)=_{\textbf{T}}(a\cdot b)\cdot c$

Standard, as sequences separated by $``,"$ are associative.

(K6) $a\cdot 1=_{\textbf{T}}1\cdot a=_{\textbf{T}}a$

Straightforward from (1I),(I1) and the the $\bullet$-rules.

(K7) $0\cdot a=_{\textbf{T}}a\cdot 0=0$

Clear from (0I).

(K8) $a\cdot(b+c)=_{\textbf{T}}a\cdot b+a\cdot c$

$a(b\vee c)\leq_\textbf{T}(a b)\vee (ac)$
follows from (D);
$(ab)\vee (ac)\leq_\textbf{T}a(b\vee c)$
follows from $a c\leq_\textbf{T}a(b\vee c)$,
$a b\leq_\textbf{T}a(b\vee c)$ and 
($\vee$I).

(K9) $(b+c)\cdot a=_{\textbf{T}}b\cdot a+c\cdot a$

similar.
\\

(K10) $1+aa^*\leq_{\textbf{T}} a^*$

We have to show that
$1\vee (a\bullet a?)\leq_\textbf{T} a?$. Here the derivation:

$\infer{1\vee a\bullet a?\vdash a?}
{\infer{1\vdash a?}
{\infer{\vdash a?}{}} &
\infer{a\bullet a?\vdash a?}{\infer{a,a?\vdash a? }{a\vdash a & a?\vdash a?}}}$
\\

(K11) $1+a^*a\leq_{\textbf{T}} a^*$

similar. 
\\

(K12) If $b+ac\leq_{\textbf{T}} c$, then $a^*b\leq_{\textbf{T}} c$

Assume we
have $\Vdash_{\KL}b\vee (a\bullet c)\vdash c$. 
We first show that in this case, 
we must have $b\vdash c$, $a\bullet c\vdash c$.
If the last rule of the derivation tree concerned the left-hand side, there is
no choice: the rule was ($\vee$I), and the claim follows,
because otherwise application was not legitimate.

So assume the last rule concerned the right-hand side;
in that case, there are several possibilities.
(I$\bullet$) can be excluded for the structure of the antecedent 
(it does not
contain $``,"$); 
the $?$-rules  change the right hand side only in case of
empty antecedent, so they can be excluded; so the only
candidates are  (D) and (I$\vee$1),(I$\vee$2). All of these rules have
a very nice property: they are (1) unary, and (2) applicable
regardless of the properties of left hand side of the
antecedent. So we can safely assume that
$b\vee (a\bullet c)\vdash c$ has been derived from 
$b\vee (a\bullet c)\vdash c'$, such that the last rule applied to
derive
$b\vee (a\bullet c)\vdash c'$ was ($\vee$I), and 
$b\vee (a\bullet c)\vdash c$ has been subsequently derived 
by applications of (D), (I$\vee$1) and (I$\vee$2) only. Consequently,
we must have two valid derivations of 
$b\vdash c'$, and $a\bullet c\vdash c'$. Now, as the rules
(D), (I$\vee$1), (I$\vee$2) do not care for the left hand side, we can
thus also derive $b\vdash c$, and $a\bullet c\vdash c$.

We can apply exactly the same argument to show that if we can derive
$a\bullet c\vdash c$, we can also derive $a,c\vdash c$. 
Thus we know
that if $\Vdash_\KL b\vee (a\bullet c)\vdash c$ holds, then we 
also have $\Vdash_\KL a,c\vdash c$ and $\Vdash b\vdash c$.
We can thus apply (?I1) to derive

$\infer{a?b\vdash c}{a,c\vdash c & b\vdash c}$
\\

(K13) If $b+ca\leq_{\textbf{T}} c$, then $ba^*\leq_{\textbf{T}} c$.

similar.
\\

This shows that $\textbf{T}$ is a Kleene algebra and completes the proof
of completeness.
\proofend

Note that for the completeness part of the proof, the cut-rule is
not needed, in fact it is not even mentioned! There are a number 
of important consequences which follow from this
result. But before we come to these, we first introduce the logic $\KL^+$
and prove a similar result.

\section{$\KL^+$}

We now present a new logic $\KL^+$, which is a fragment of $\KL$.
Our motivation is the following: we would like to intuitively interpret 
$\alpha?$ as a sort of progressive aspect of the event
$\alpha$, that is: if $\alpha$ means ``(in some interval) $\alpha$ is completed",
then $\alpha?$ should mean: ``(in some interval) $\alpha$ 
is going on". Obviously, then (ax?) is way too strong: we
cannot assert that just anything is going on. What we rather can assert
is the weaker implication: if something happens (possibly several times)
in an interval, then it is happening in this interval:
if in some interval I ate a pizza, I was eating a pizza in this interval,
though not the converse, for my starting and finishing
the pizza might be laying outside the interval.\footnote{This is known as 
the imperfective paradox, 
see \cite{moenssteedman:ontology}; if I crossed the street, I was
crossing it, but not vice versa.} In terms of logical consequence, 
the progressive $\alpha?$ of an atomic
event $\alpha$ can be characterized as follows: it follows from any
(non-zero) number of iterations (in terms of $\cdot$) 
of $\alpha$ and $\alpha?$, 
and from nothing else, except for transitivity and logical laws which govern the
other connectives.\footnote{Of course, this is a gross
simplification; linguistically speaking there is much more to it.
See for example \cite{galton:aspect}.}
We devise $\KL^+$ in order
to agree with our intuition on the progressive aspect of events;
one of our main results will be that whereas the $?$ of $\KL$ can be
interpreted as Kleene star, the $?$ of $\KL^+$ can be interpreted
as Kleene plus.

As we said, the syntax of $\KL^+$ is identical to $\KL$. 
Regarding its consequence relation and sequent calculus, we
can re-use also most of the rules and axioms of $\KL$. So we just
say which rules are discarded, and which ones are new.
To obtain the rules for $\KL^+$, we take away 
the axiom (ax?)
and substitute it with the weaker axiom  (+?)
It is clear that (+?) is derivable in $\KL$.
Note that (+?) \textit{cannot} derive a sequent of the form
$\vdash\alpha?$ or
$\alpha\vdash\alpha\alpha?$. This is our intention; but as a consequence
we also need to reconsider the rules (?I1) and (?I2), which we 
replace by (?+1),(?+2): 
\\

(+?) $\infer{\Gamma\vdash\alpha?}{\Gamma\vdash \alpha}$
$\qquad$ (?+1) $\infer{\alpha?,\Gamma\vdash\beta}
{\alpha,\beta\vdash\beta & \alpha,\Gamma\vdash\beta}$
$\qquad$ (?+2) $\infer{\Gamma,\alpha?\vdash\beta}
{\beta,\alpha\vdash\beta & \Gamma,\alpha\vdash\beta}$
\\

Regarding these rules, we have to say the following: given 
the cut rule, (?I1) and (?I2) are derivable from (?+1) and (?+2),
respectively (just assume you have the premises of (?I1), 
$\alpha,\beta\vdash\beta$, $\Gamma\vdash\beta$; by cut, you get
$\alpha,\Gamma\vdash\beta$). The converse is not true: 
we cannot derive (?+1) and (?+2) without (ax?). In particular, it is easy to 
show that without these rules, we cannot derive the sequent
$\alpha?,\alpha\vdash\alpha\bullet(\alpha?)$: just ask what was the
last rule applied to this sequent: the 
(I$\bullet$) rule was not applicable to derive this sequent, 
for $\alpha?\not\vdash\alpha$, and any other rule is out of 
the question for the syntactic form of the sequent.
As long as we have the cut rule, we can moreover
derive (I?1),(I?2). What if the cut rule is lacking? I
do not see how to derive the two, but from 
cut admissibility, which we prove later on
for $\KL,\KL^+$, it follows that the two do not allow
us to derive anything we could not derive without them in
 $\KL^+$. 
On a related note, note that
(?+1),(?+2) are also admissible in $\KL$, that is, they would not
add anything new to the calculus, though I do not see how they 
can be derived. As we have the cut rule, we could thus
substitute (I?1),(I?2) by (?+1),(?+2) in $\KL$, thereby 
giving a more uniform treatment of $\KL$ and $\KL^+$. The 
reason we have not chosen this presentation is the 
following: for the alternative
presentation, (as far as I can see) 
we need the cut-rule to prove that its term-algebra is a Kleene-algebra,
so our simple semantic proof of cut admissibility (see section 9)
for ``standard" $\KL$ would no longer work.



It is clear that $\KL^+$ is a fragment 
of $\KL$, as its axioms and inference rules
are derivable in $\KL$. The question is: what exactly is the expressive
power of $\KL^+$? We will show the following: we can give it
a strongly complete semantics
in terms of (slightly modified) Kleene algebras and in terms of language models,
by only a minor change in interpretation: we interpret 
the connector $?$ as Kleene \textit{plus} instead of star. So
what changes with the new axiom is essentially the meaning of $?$,
and nothing else. The proof of this is however slightly more
complicated, as we have to make several steps consisting in algebraic
embeddings. It is well known that we can 
define $a^+$ as $aa^*$ (or $a^*a$). But nonetheless we cannot
work directly with Kleene algebras, putting 
$\osigma(\alpha?)=\osigma(\alpha)^*\osigma(\alpha)$, because we have
$a^*a=_\KA aa^*$, but none of the corresponding sequents is provable,
and thus completeness fails.

As we do not work directly with
Kleene algebras, we have to work with a variant and two embeddings.
We interpret $\KL^+$ in a class of algebras
which we call $\KA^\#$. We show strong completeness for this 
semantics, where we first go
the usual way: we show that the term 
algebra $\textbf{T}^+$ of $\KL^+$ is a 
$\KA^\#$ algebra, such that if $\not\Vdash_{\KL^+}\Gamma\vdash\alpha$, 
then $\not\models_{\KA^\#}\Gamma\vdash\alpha$. 

We then have to show that strong completeness
also holds for the language-theoretic semantics. To this aim,
we devise two maps $i,j$, which map $\KA$ terms onto $\KA^\#$ terms
and vice versa. We show some validity-preserving 
properties of these maps,
which allow us to extend language-completeness results from
$\KA$ to $\KA^\#$, without having to perform a complicated proof from scratch
as in \cite{kozen:completeness}.

\section{Algebraic Semantics: $\KA^\#$-algebras}

We now define a variant of Kleene algebras, namely $\KA^\#$ or
Kleene plus algebras. We have the connectives $+,\cdot,\#$, where
$\#$ is unary, and constants 0,1.  We have (K1)--(K9) as in $\KA$;
then things change. We list the new axioms (K14+)--(K17+), together
with the more expectable, but ``wrong" axioms (K10+)--(K13+), just to show
how the latter can be derived from the former, but not vice versa:
\\

(K14+) $a+aa^\#\leq a^\#$. 

(K15+) $a+a^\#a\leq a^\#$.

(K16+) If $ab+ac\leq c$, then $a^\#b\leq c$.

(K17+) If $ba+ca\leq c$, then $ba^\#\leq c$.
\\

($\quad$(K10+) $1+a(1+a^\#)\leq 1+a^\#\quad$)

($\quad$(K11+) $1+(1+a^\#)a\leq 1+a^\#\quad$)

($\quad$(K12+) If $b+ac\leq c$, then $((1+a^\#)b)=b+a^\#b\leq c\quad$) 

($\quad$(K13+) If $b+ca\leq c$, then $(b(1+a^\#))=b+ba^\#\leq c\quad$) 
\\

(K10+)--(K13+) are very ``expectable": they just consist in
(K10)--(K13), where each time, $a^*$ is replaced by $1+a^\#$.
If we ``read" $a^\#$ as $aa^*$ (or $a^*a$), then they are valid
in $\KA$,
because then we have $a^*=_\KA 1+a^\#$. But the reader has
to keep in mind that we have a different algebra here, where
$*$ does not exist as a connective. (K14+) and (K15+) seem to be redundant
with (K10+),(K11+), and in fact they create redundancy: we can derive
(K10+) from (K14+) and (K11+) from (K15+), but not vice versa
(at least I do not see how): we can easily derive  
$1+a+aa^\#\leq 1+a^\#$ from (K10+); but we still have to get rid of the 1.
Conversely, we can derive (K10+),(K11+) from (K14+),(K15+):
if $aa^\#\leq a^\#$, then $aa^\#\leq 1+a^\#$; this means:
$aa^\#+1+a^\#=1+aa^\#+1+a^\#=1+a^\#$
iff $1+aa^\#\leq 1+a^\#$.
Same holds for the pairs (K12+),(K13+) and (K16+),(K17+):
we can derive (K12+) from (K16+), because if $b+ac\leq c$, then
by strong transitivity $ab\leq c$, and thus $a^\#b\leq c$. 
The converse does not hold, 
and in particular, (K12+),(K13+) do not allow us to derive $a^\#\leq a+a^\#a$.
With (K14+),(K16+) it is an easy exercise (put $b:=1,a:=a,c:=a+a^\#a$).
So we axiomatize $\KA^\#$ by (K1)--(K9),(K14+)--(K17+);
we leave the other axioms for illustration and because they turn
out to be useful for proving properties of our later embeddings.

Most basic properties of $\KA$ transfer to $\KA^\#$: $\leq_{\KA^\#}$
is defined over $+$ in the usual fashion and is thus reflexive,
transitive and antisymmetric. 
The same holds for the fact that $\leq_{\KA^\#}$ 
respects concatenation, because of distributivity, and we thus have
strong transitivity. 
We now devise an \textbf{interpretation} of $\KL^+$ in $\KA^\#$. 
We define $\sigma$ as usual, and $\osigma$ as before, the expectable
exception that
\\

$\osigma(\alpha?):=\osigma(\alpha)^\#$.
\\

\section{Completeness of the Algebraic Semantics}

We take the usual definitions, and prove soundness and completeness
of $\KL^+$ for $\KA^\#$:

\begin{thm}
We have $\models_{\KA^\#}\Gamma\vdash\alpha$  if and only if
$\Vdash_{\KL^+}\Gamma\vdash\alpha$.
\end{thm}

\proofbeg
We start with the \textit{if} direction (\textbf{\textit{soundness}}). 
Most of the axioms can be skipped, as the old soundness arguments
remain valid (as $\KL^+$ is a fragment of
$\KL$, and most equations of $\KA$ are valid in $\KA^\#$). 
We only need to prove that the inference rules regarding
$?$ are sound with respect to $\KA^\#$.
\\

(+?) 
We have to show that if $\osigma(\Gamma)\leq_{\KA^\#} \osigma(\alpha)$, 
then $\osigma(\Gamma)\leq_{\KA^\#} \osigma(\alpha)^\#$.
That is straightforward with (K14+),  according to which $a\leq_{\KA^\#} a^\#$,
and transitivity of $\leq_{\KA^\#}$.
\\

(?+1) 
We make the usual induction. Assume the premises
are satisfied, that is, we have
$\models_{\KA^\#}\alpha,\Gamma\vdash\beta$ and
$\models_{\KA^\#}\alpha,\beta\vdash\beta$; consequently,
$\osigma(\alpha)\osigma(\Gamma)\leq_{\KA^\#} \osigma(\beta)$ 
and $\osigma(\alpha)\osigma(\beta)\leq_{\KA^\#} \osigma(\beta)$. 
Consequently, by the order definition,
we have $\osigma(\alpha)\osigma(\Gamma)+
\osigma(\alpha)\osigma(\beta)\leq_{\KA^\#} \osigma(\beta)$.  
By (K16+) it follows that 
$\osigma(\alpha)^\#\osigma(\Gamma)\leq_{\KA^\#} \osigma(\beta)$,
and thus
$\models_{\KA^\#}\alpha?,\Gamma\vdash\beta$.
\\

(?+2) 
similar.
\\

(I?1) 
Assume the premises hold. Then we have 
$\osigma(\Delta)\leq_{\KA^\#} \osigma(\alpha)$ and 
$\osigma(\Gamma)\leq_{\KA^\#} \osigma(\alpha)^\#$. We have to show that
$\osigma(\Delta)\osigma(\Gamma)\leq_{\KA^\#} \osigma(\alpha)^\#$ . 
Because $\leq_{\KA^\#}$ respects concatenation, we know that
$\osigma(\Delta)\osigma(\Gamma)\leq_{\KA^\#} \osigma(\alpha)\osigma(\alpha)^\#$. 
From $(K14+)$ and transitivity it follows that
$\osigma(\Delta)\osigma(\Gamma)\leq_{\KA^\#} \osigma(\alpha)^\#$.
\\

(I?2) 
similar.
\\

This completes the soundness direction. 
\\

\textbf{\textit{Completeness}}.

We do the same construction as before, now call $\textbf{T}^+$ the
algebra $(T,\vee,\bullet,?,0,1)$; we define $T$ to be 
the set of $\textit{Form}(\textit{Var})/\negmedspace\negmedspace\sim_{\KL^+}$, that is, the set
of congruence classes of $\KL^+$ formulas under the congruence
$\sim_{\KL^+}$, which is logical equivalence in $\KL^+$.
We put $a\leq_{\textbf{T}^+}b$ iff $\Vdash_{\KL^+}\alpha\vdash \beta$
for some $\alpha\in a,\beta\in b$, 
and put $a=_{\textbf{T}^+}b$ iff $a\leq_{\textbf{T}^+}b$ 
and $b\leq_{\textbf{T}^+}a$.
We skip the proof that $\sim_{\KL^+}$ is a congruence
over connectives, which is straightforward, and write 
congruence classes as if they were formulas.

To show completeness, we have to show that $\textbf{T}^+$
is a $\KA^\#$-algebra. 
Again, we can skip  most of the equations as the proof is
identical to the one for $\KL$;
we check only those in which $\#$ (which is $?$ in $\textbf{T}^+$)
occurs, and which are not derivable. 
\\

(K14+) $a+aa^\#\leq_{\textbf{T}^+} a^\#$. 

It is obvious that $a\leq_{\textbf{T}^+} a?$; we need to show that 
$aa?\leq_{\textbf{T}^+} a?$. That is also easy:

$\infer{aa?\vdash a?}{a\vdash a & a?\vdash a?}$; 

and consequently
$a\vee aa?\leq_{\textbf{T}^+} a?$.
\\

(K15+) similar.
\\

(K16+) If $ab+ac\leq_{\textbf{T}^+} c$, then $a^\#b\leq_{\textbf{T}^+} c$. 

Assume we have $(ab)\vee (ac)\leq_{\textbf{T}^+} c$. Then by the same
argument as for $\KL$, we can conclude that we have both
$\Vdash_{\KL^+}a,b\vdash c$ and $\Vdash_{\KL^+}a,c\vdash c$. 
 Then by (?+1), we derive

$\infer{a?b\vdash c}
{ab\vdash c & ac\vdash c}$
\\

(K17+) is similar.
\\

This completes the proof.
\proofend

Note again that the cut rule is not needed for the completeness proof.
But this is not as satisfying as the other completeness theorem,
because we do not know nearly as much about $\KA^\#$ algebras
as about $\KA$ algebras. We now establish some results showing
the algebraic correlation between $\KA$ and $\KA^\#$.

\section{Algebraic Relations between $\KA$ and $\KA^\#$}

We now prove  three lemmas, which together give us a very 
powerful result. Let $(K,+,\cdot,*,0,1)$ be a $\KA$ algebra,
$(K',+,\cdot,\#,0,1)$ be a $\KA^\#$ algebra. We define the
map $i$ as follows: 
\begin{enumerate}
\item for $a$ atomic, $i(a)=a$;
\item $i(a+b)=i(a)+i(b)$;
\item $i(a\cdot b)=i(a)\cdot i(b)$;
\item $i(a^*)=1+(i(a))^\#$.
\end{enumerate}

$i$ is a map from $\KA$-terms to $\KA^\#$-terms. 
Its most important property is the following, which is
intuitively clear, but still needs to be proved.

\begin{lem}
If $a\leq_\KA b$, then $i(a)\leq_{\KA^\#}i(b)$.\footnote{The inverse
implication can also be proved in much the same way as lemma 4, just using
lemma 5 instead of lemma 3. This result does however not play a 
role in the sequel, so we do not explicitly state it.}
\end{lem}

\proofbeg
We have $a^*=_\KA 1+aa^*=_\KA 1+a^*a$.
Consider a class of algebras $\KA^{\#'}$, which consists of all
algebras which are $i$-images of $\KA$-algebras. 
Its axioms are exactly the $i$-images of
axioms (K1+)--(K13+), but the terms have a different syntax:
$(1+a^\#)$ is just a single constructor over $a$, to which the distributive,
commutative, associative laws etc. do not apply. 
It is immediately clear
(by the purely syntactic translation) that 
$a\leq_\KA b$ if and only if $i(a)\leq_{\KA^{\#'}}i(b)$.
Now, as we have shown before, all axioms (K1+)--(K13+)
are valid in $\KA^{\#}$. Moreover, the syntactic difference
of $\KA^\#$ and $\KA^{\#'}$ needs not bother us, as long as we are
trying to prove that $\KA^{\#'}$ inequations are valid in
$\KA^{\#}$: in the latter, nothing prevents us from treating
$(1+a^\#)$ as a unit, that is, not applying any of distributive,
commutative or associative laws to it. So the inequations valid
in $\KA^{\#'}$ are a (proper) subset of the inequations valid in
$\KA^{\#}$; and so, as $i(a)\leq_{\KA^{\#'}}i(b)$, we have
$i(a)\leq_{\KA^{^\#}}i(b)$.
\proofend

Now we define a sort of inverse of $i$; define the map
$j$ from $\KA^\#$ to $\KA$ terms as follows:
\begin{enumerate}
\item $j(a)=a$ for atomic $a$;
\item $j(a+b)=j(a)+j(b)$;
\item $j(a\cdot b)=j(a)\cdot j(b)$;
\item $j(a^\#)=j(a)(j(a))^*$.
\end{enumerate}


\begin{lem}
If $j(a)\leq_\KA j(b)$, then $a\leq_{\KA^\#}b$.
\end{lem}

\proofbeg
Assume we have $j(a)\leq_\KA j(b)$. As $j$ is trivial for all
connectives but $\#$, we only treat terms of the form $a^\#$. 
It follows from
the the map $j$ that in $j(a),j(b)$ all occurrences $a^*$ occur in 
subterms $(aa^*)$. Now form the images $i(j(a)),i(j(b))$, which are
$\KA^\#$-terms. By lemma 3, we have
$i(j(a))\leq_{\KA^\#}i(j(b))$. Here, all occurrences of $\#$ are in subterms
$(a(1+a^\#))$. By the usual laws, we have
$(a(1+a^\#))=_{\KA^\#}(a1+aa^\#)=_{\KA^\#}(a+aa^\#)$. Moreover, we have 
$a^\#=_{\KA^\#}a+aa^\#$; so we have $i(j(a))=_{\KA^\#}j^{-1}(j(a))=_{\KA^\#}a$; 
same for
$b$; and thus $a\leq_{\KA^\#}b$.
\proofend

\begin{lem}
If $a\leq_{\KA^\#}b$, then $j(a)\leq_\KA j(b)$.
\end{lem}

\proofbeg
We show this in the same fashion as lemma 3: construct an
intermediate class of algebras $\KA'$, where $aa^*$ is a single constructor over
$a$, to which associative, distributive laws do not apply, and
where all terms
have the form $j(a)$ for some $\KA^\#$-term $a$, and all $\KA^\#$-axioms 
in their translation under $j$
are valid. We show that all these $j$-images of $\KA^\#$-axioms 
are also valid (under the different syntactic reading
of $\KA$!) within $\KA$: 

(j(K14+)) $a+aaa^*\leq aa^*$. 

That is obviously derivable by $1\leq_\KA a^*$, $a\leq_\KA a^*$, $a\leq_\KA a$,
and strong transitivity.
(K15+) is similar. 

(j(K16+)) If $ab+ac\leq c$, then $aa^*b\leq c$. 

Just put $b:=ab$ in the quasi-inequation 
``if $b+ac\leq_\KA c$, then $a^*b\leq_\KA c$", and
we obtain $a^*ab\leq c$ from the premise. As we have 
$aa^*=_{\KA} a^*a$, the consequence is valid in $\KA$ if the
premises are.
(K17+) is similar.

Again, all axioms of $\KA'$
are valid in $\KA$; the possible manipulations of $\KA'$ are
also a subset thereof, so everything valid in $\KA'$ is valid in $\KA$;
if $a\leq_{\KA^\#}b$, then $j(a)\leq_{\KA'} j(b)$, and thus $j(a)\leq_\KA j(b)$.
\proofend

These results are of course of some value on their own, because they
show us how the intuitive correlation of $\KA$ and $\KA^\#$ corresponds 
to formal notions. Their importance for $\KL^+$ reveals itself in the
next section, where we consider language models.

\section{Completeness of Language-theoretic Semantics}

As is well known, we can interpret $\KA$-terms as regular
expressions; atomic terms are interpreted as letters (or as 
languages),
$+$ as union, $\cdot$ as concatenation, and $*$ as Kleene
star, that is, union of all finite iterations. Let $a$ be
a $\KA$ term; we denote the language it denotes under this
interpretation by $\Vert a\Vert$.
If $a$ is a term over the set of (atomic) generators $A$, 
then $\Vert a\Vert\subseteq A^*$ (the set of all finite strings of
$A$-symbols). The following 
fundamental theorem for Kleene algebras was proved by 
Kozen in \cite{kozen:completeness}:

\begin{thm}
(Kozen)
We have $a\leq_\KA b$ if and only if $\Vert a\Vert\subseteq\Vert b\Vert$.
\end{thm}

From this theorem and the lemmas of the preceding section we can easily
derive a similar result for $\KA^\#$ algebras. Let $a$ be a $\KA^\#$
term. We can interpret $\KA^\#$ terms as languages as follows:
\begin{enumerate}
\item $\Vert a\Vert^\#=\{a\}$, for atomic $a$;
\item $\Vert a+b\Vert^\#=\Vert a\Vert^\#\cup \Vert b\Vert^\#$;
\item $\Vert a\cdot b\Vert^\#=\Vert a\Vert^\#\cdot \Vert b\Vert^\#$;
\item $\Vert a^\#\Vert^\#=\Vert a\Vert^\#\bigcup_{n\in\mathbb{N}_0}(\Vert a\Vert^\#)^n$.
\end{enumerate}
Thus the $\#$ is interpreted as Kleene plus. We now show the following:

\begin{thm}
We have $a\leq_{\KA^\#}b$ if and only if 
$\Vert a\Vert^\#\subseteq\Vert b\Vert^\#$.
\end{thm}

\proofbeg
\textit{If}: Assume $\Vert a\Vert^\#\subseteq\Vert b\Vert^\#$. 
We can read
$a,b$ as regular expressions with Kleene plus. By definition of the
Kleene plus in terms of the star, we have $\Vert a\Vert^\#=\Vert j(a)\Vert$,
same for $b$. Consequently, $\Vert j(a)\Vert\subseteq\Vert j(b)\Vert$,
and by theorem 6, $j(a)\leq_\KA j(b)$. By lemma 4,
we obtain $a\leq_{\KA^\#}b$. 

\textit{Only if}: Assume $a\leq_{\KA^\#}b$. 
By lemma 5, we then have $j(a)\leq_\KA j(b)$, and thus
$\Vert j(a)\Vert\subseteq \Vert j(b)\Vert$. So the claim
follows the fact that $\Vert j(c)\Vert=\Vert c\Vert^\#$.
\proofend

From this follows that both $\KL$ and $\KL^+$ have a complete
language-theoretic semantics:

\begin{thm}
We have 
\begin{enumerate}
\item $\Vdash_\KL\Gamma\vdash\alpha$ if and only if 
$\Vert\osigma(\Gamma)\Vert\subseteq \Vert\osigma(\alpha)\Vert$; and
\item $\Vdash_{\KL^+}\Gamma\vdash\alpha$ if and only if 
$\Vert\osigma(\Gamma)\Vert^\#\subseteq \Vert\osigma(\alpha)\Vert^\#$.
\end{enumerate}
\end{thm}

Now, as under this interpretation, formulas of both $\KL$ and
$\KL^+$ denote regular languages, and the problem whether one
regular language (represented, e.g., as a regular expression)
is a subset of another one is decidable, we immediately get the following:

\begin{thm}
$\KL,\KL^+$ are decidable, that is, for any sequent $\Gamma\vdash\alpha$,
we can effectively decide whether $\Vdash_{\KL}\Gamma\vdash\alpha$,
$\Vdash_{\KL+}\Gamma\vdash\alpha$ hold.
\end{thm}

So in order to decide whether a sequent is valid, we can just go
over the language-theoretic interpretation of formulas as regular
expressions. A more direct way to establish the
decidability of $\KL,\KL^+$ is by showing that for every derivable
sequent, there is a proof which does not make use of the cut rule.
%
%

\section{Cut Admissibility}

We now show that in $\KL,\KL^+$ cut is admissible, 
that is, for every proof of a sequent in the two calculi
there is a  cut-free proof.
\footnote{Contrary to some other
usage, we do not speak of cut elimination, because we only show
the existence of a cut free proof, whereas cut elimination means
that from a proof using cut we can effectively construct a cut free
proof.} 
This is important for the following reason:
the cut rule is the only rule in which there is material
in the antecedents, which is not in the consequent; so when we 
want to check whether a sequent is derivable in our
calculi, it is the only rule which makes the search space infinite;
for all other rules, we know how the antecedents have
to look like (in the sense of: there is a finite number of 
possible choices). So from cut admissibility follows that $\KL,\KL^+$ are
decidable also ``inside the calculus", without a detour over
semantics. 

How does this result follow? We established the
soundness of the cut rule. But in order to show the completeness
direction of the algebraic semantics, proving the term 
algebra of $\KL,\KL^+$ is a $\KA,\KA^\#$ algebra, respectively, 
we did not make any use of the cut rule. 
Now let $\KL_{\mathit{cf}},\KL^+_{\mathit{cf}}$ be the logics which 
result from taking all axioms and inference 
rules of $\KL$ and $\KL^+$, respectively,
except for the cut rule. To prove their soundness and completeness for 
$\KA,\KA^\#$, we can take over the proofs for $\KL,\KL^+$ without
any change, except that we can do away with soundness of cut. 
So we have $\Vdash_{\KL_{\mathit{cf}}}\Gamma\vdash\alpha$ 
iff $\models_{\KA}\Gamma\vdash\alpha$ iff $\Vdash_{\KL}\Gamma\vdash\alpha$,
same for $\KL^+$ and $\KL^+_{\mathit{cf}}$. The only problem with this reasoning
is that the interpretation $\osigma$ is a homomorphism rather than a
bijection, as it maps both $``,"$ and $``\bullet"$ to $``\cdot"$. So
in addition we need a proof of the fact that 
$\Vdash_{\KL}\gamma_1,...,\gamma_i\vdash\alpha$ iff 
$\Vdash_{\KL}\gamma_1\bullet...\bullet\gamma_i\vdash\alpha$, and the
same for $\KL_{\mathit{cf}}$. This is however easy to show (cf. \cite{galatos:residuated},
Proposition 7.1). From this follows: 

\begin{thm}
In both $\KL,\KL^+$, the cut rule is admissible.
\end{thm}

\section{Conclusion}

We have presented two propositional substructural logics,
strongly inspired by Kleene algebras and 
some considerations on processes. We have proved strong completeness 
theorems for algebraic semantics as well
as for language models. We have also
proved cut admissibility, from which follows the decidability of the
calculus by purely syntactic means (though the cut admissibility
proof itself is semantic). 
By our strong completeness, proof search in the calculus of $\KL$
can be reduced to the validity of an inequation in all Kleene algebras
(PSPACE-complete). There remains
the question whether for some subclass 
of formulas, we can do substantially better
(this is an approach commonly taken, see \cite{hovland:inclusion}): 
there might fall off a good 
algorithm for checking the inclusion relation
of languages denoted by two regular expressions.
Another question is the following: as can be seen from already
existing work (see e.g. \cite{buszkowski:action}), if we enrich
a substructural logic having an exponential as $?$ with implication
(or vice versa), undecidability strikes rather quickly. Still,
the decidability results for $\KL,\KL^+$ seem robust;	 
in particular, 
we conjecture that the \textit{external} consequence
relations (the smallest structural consequence relations generated
by the two, see e.g. \cite{galatos:residuated}) 
of the two logics are decidable.
These seem to us the most natural and
interesting questions to ask. 

\bibliographystyle{eptcs}
\bibliography{alles_7_14}
\end{document}